\documentclass{ws-ijmpa}
\usepackage[super,compress]{cite}
\usepackage{graphicx}
\begin{document}
\markboth{D. MOMENI et al.}
{Tolman-Oppenheimer-Volkoff equations in non-local
$f(R)$ gravity}

%
\catchline{}{}{}{}{}
%

\title{Tolman-Oppenheimer-Volkoff equations in non-local
$f(R)$ gravity
}

\author{Davood Momeni}

\address{Eurasian International Center for Theoretical Physics and Department of General
\& Theoretical Physics, Eurasian National University, Astana 010008, Kazakhstan\\
d.momeni@yahoo.com}

\author{H. Gholizade}

\address{Department of Physics, Tampere University of Technology\\  P.O.Box 692, FI-33101 Tampere, Finland\\
hosein.gholizade@gmail.com}
\author{Muhammad Raza}
\address{Department of Mathematics, COMSATS Institute of
Information Technology, Sahiwal 57000, Pakistan\\
State Key Lab of Modern Optical Instrumentation, Centre
for Optical and Electromagnetic Research, Department of Optical
Engineering, Zhejiang University, Hangzhou 310058, China\\
mraza@zju.edu.cn}
\author{Ratbay Myrzakulov}
\address{Eurasian International Center for Theoretical Physics and Department of General
\& Theoretical Physics, Eurasian National University, Astana 010008, Kazakhstan\\
rmyrzakulov@gmail.com}

\maketitle

\begin{history}
\received{Day Month Year}
\revised{Day Month Year}
\end{history}

\begin{abstract}
Non-local $f(R)$ gravity was proposed as a powerfull alternative to
general relativity (GR) .
This theory  has potentially adverse implications for infrared (IR)
regime as well as ultraviolent(UV) early epochs. However, there are
a lot of powerful features, making it really user-friendly. A
scalar-tensor frame comprising two auxiliary scalar fields, used to
reduce complex action. However this is not the case for the
modification complex which plays a distinct role in modified
theories for gravity. In this work, we study the dynamics of a
static, spherically symmetric object. The interior region of
spacetime had rapidly filled the perfect fluid. However, it is
possible to derive a physically based model which relates interior
metric to non-local $f(R)$.  The Tolman-Oppenheimer-Volkoff (TOV)
equations would be a set of first order differential equations from
which we can deduce all mathematical (physical) truths and derive
all dynamical objects. This  set of dynamical equations govern
pressure $p$, density $\rho$, mass $m$ and auxiliary fields
$\{\psi,\xi\}$.  The full conditional solutions are evaluated and
inverted numerically to obtain exact forms of the compact stars Her
X-1, SAX J 1808.4-3658 and 4U 1820-30 for non-local Starobinsky model of
$f(\Box^{-1}R)=\Box^{-1}R+\alpha \Big(\Box^{-1}R\Big)^2$.  The program solves the differential equations
numerically using adaptive Gaussian quadrature. An ascription of
correctness is supposed to be an empirical equation of state
$\frac{P}{P_c}= a (1- e^{-b\frac{\rho}{\rho_c}})$ for star which is
informative in so far as it excludes an alternative non local
approach to compact star formation. This model is most suited for
astrophysical observation.
\keywords{Higher-dimensional gravity and other theories of gravity;
 neutron stars;
 thermodynamic processes; conduction; convection; equations of state.}
\end{abstract}

\ccode{PACS numbers: 04.50.-h,97.60.Jd,95.30.Tg}

\section{Introduction}
Modern cosmology is based on the relativistic description of large
scale Universe.  The situation is made worse when the Universe
contains expansion and so accelerating. We are restricted to obey
different observational data in favor of an accelerating universe at
very large scales \cite{obs1}-\cite{obs3}. Accelerating expansion is
largely responsible for the fact that each galaxy in large scale are
expanding the capacity to expand themselves. They provide evidence
of the accelerating expansion of Universe and hence for the
existence of the mysterious dark energy which drives this process.
There is no doubt that these data are modifying the classical
description of gravitation, general relativity (GR). We will engage
other theories in modifying our GR to finally understand this epoch.
Universe  may be accelerating due to matter and geometry of the full
action. Two contrasting methods for accelerating scenario have been
identified. At the first candidate, an exotic matter fluid was used
in the full action of theory to realize a variety of accelerating
and decelerating phases.  They had to prove that GR respond to
cosmological needs in a world of accelerating expansion that we
observed earlier. At the next stage, a geometrical modification was
used in the action to study a variety of accelerating and
decelerating epochs
 \cite{Nojiri:2010wj}-\cite{ijgmmp}).
As individuals, there are things we can perform to improve the
situation, firstly as geometrical modifications and secondly as
fluids.\par Firstly there is the belief that the  Buchdahl was the
founder of the $f(R)$ modified gravity  \cite{Buchdahl}. A model is
proposed to explore the viable range of parameters of cosmological
background due to replacing the Ricci scalar $R$ by an arbitrary
function $f(R)$. Proposed action by nonlinear higher terms should be
applied to the large scale. Among the proposed models it has been
considered that Universe expansion caused in the late time where low
curvature corrections were formed as space time were filled. Similar
difficulties may apply to the higher order corrections
 as Gauss-Bonnet (GB) term, $G=R^2-4R_{\mu\nu}R^{\mu\nu}+R_{\mu\nu\lambda\sigma}
R^{\mu\nu\lambda\sigma}$.
Indeed, in four dimensional Einstein-Hilbert action, these GB terms
have no contribution if they appear in non-minimally coupled form.
The reason backs to the topology and topological invariants.
Coupling of GB term to matter fields provides more dynamical
features than before. In comparison to $f(R)$, the modified GB
gravity was proposed to be the $f(G)$ gravity as model for dark
energy in \cite{sergei3}. Several interesting cosmological features
of this type modified gravity were studied
\cite{Capozziello:2014bqa}-\cite{Nojiri:2006je}. One of the most
important motivations of GB gravity is that it arises naturally from
string theory. Also, it appears as a non-minimally coupling term to
higher order scalar fields \cite{Horndeski}. The list of possible
modifications is not limited upto the above mentioned models. The
non-local models are also so interesting  theories and enough
popular as modified gravities. This last case, is studied in our
paper. We'll present a short review of the subject in Sec.
(\ref{Non-local $f(R)$ gravity}).   \par These mass densities that
are localized then immediately collapse, releasing thermodynamics
that can cause a hydrodynamic pressure impose on the mass
distribution. Following the collapse of the matter distribution,
mass transformed itself into a compact star. The compact star has a
finite size and a huge amount of mass \cite{book}. It was an open
competition between different massive objects that enabled the
neutron star carriers the opportunity to exist in the space time.
The neutron star has very strong surface gravity and thermodynamics.
We can detect these massive objects by looking at the Doppler shift
in spectral lines emitted by atoms in the surface. The major
elements mass $M\sim1.4 M_{\odot}$ and radius $R\sim 10 Km$ of
inputs are used to control the whole dynamics. Specific surface
gravity known as the relative surface gravity of the neutron star
also may be studied. Not much gravity on the earth, you can find
more on the compact star $\kappa=2\times 10^{11}$! The field of
force associated with compact star have both electric and magnetic
components and contains a definite amount of electromagnetic energy.
This means there is a greater difference between individual plants
which makes the species and neutron stars. For some basic
parameters, for example emission above-ground level, electromagnetic
effects were stronger than thermodynamic effects.

It was shown that the thermodynamic parameters, pressure $p$, energy
density $\rho$, the mass function $M$ and radius $R$ are related
according to a set of "state" equations, called as TOV equations in
GR \cite{tov1}-\cite{tov3}. It defines a set of first order
differential equations for a spherically symmetric metric which is
filled with the perfect fluid with pressure and energy density. One
main goal of our paper is to derive explicitly, the forms of TOV
equations for non-local $f(R)$ gravity. Recently TOV equations and
dynamics of stars have been investigated for different types of
modified gravity models from $f(R)$, $f(G)$ and $f(T)$ ($T$ is
torsion ) numerically and in a non-perturbative scheme
 \cite{Astashenok:2014pua}-
\cite{Astashenok:2014gda}. \par
Our plan in this letter is the following scheme: In Sec.
(\ref{Non-local $f(R)$ gravity}) we present non-local $f(R)$ gravity
as an alternative theory for gravity. In Sec. (\ref{Spherically
symmetric model of compact stars}) we derive equations of motion for
a spherically symmetric star. In Sec.
(\ref{Tolman-Oppenheimer-Volkoff Equations}) we pass to the
dimensionless parameters and we redefine all functions to obtain TOV
equations.  In Sec. (\ref{Isotropic compact star}) we study an
isotropic model of compact star using astrophysical data. We
conclude and summarize in Sec. (\ref{Summary and conclusion}). Some
preliminary formulae are presented in Sec. (\ref{APPENDIX}).

\section{Non-local $f(R)$ gravity}\label{Non-local $f(R)$ gravity}
Non-local corrections to the Einstein-Hilbert action proposed an
attempt to obtain a "healthy" version of GR with added quantum loop
corrections \cite{Deser:2007jk}-\cite{Deffayet:2009ca}. The simple
problem was how to explain the current acceleration expansion of the
Universe and to get the large numbers from inverse differential
operator(s). Indeed, this modification is refereed as an \texttt{IR
non-local modification} of General Relativity. After the original
one, another model proposed as non-local $F(R)$ gravity
\cite{Nojiri:2007uq}-\cite{Jhingan:2008ym}. It was essentially a
viable،  IR modification of the original $f(R)$ gravity. Our study
will be started from this motivated idea (for a dicussion on singularities and cosmological aspects see \cite{arXiv:1104.2692}). Let us  start by the
appropriate form of action for non-local $f(R)$ gravity:
\begin{eqnarray}
\label{nl1}
S=\int d^4 x \sqrt{-g}\left\{
\frac{1}{2\kappa^2}R\left(1 + f(\Box^{-1}R)\right) + {\cal L}_{\rm matter}
\right\}\ .
\end{eqnarray}
Here $f$ is an arbitrary function of $R$ ,
$\Box=\nabla_{\mu}\nabla^{\mu}=\frac{1}{\sqrt{-g}}\partial_{\mu}\Big(\sqrt{-g}g^{\mu\nu}\partial_{\nu}\Big)$
stands for  Laplace-Beltrami (d’Alembertian) operator in equations
of motion. We also  adopted the commonly used signature of the
metric $g_{\mu\nu}$ as $(+---)$. With this signature,  the
curvelinear derivative operator and the Riemann tensor read as the
following:
\begin{eqnarray}
&&\nabla_{\mu}V_{\nu}=\partial_{\mu}V_{\nu}-
\Gamma_{\mu\nu}^{\lambda}V_{\lambda},\\&&
R^{\sigma}_{\;\mu\nu\rho}=\partial_{\nu}\Gamma^{\sigma}_{\mu\rho}-\partial_{\rho}
\Gamma^{\sigma}_{\mu\nu}+\Gamma^{\omega}_{\mu\rho}\Gamma^{\sigma}_{\omega\nu}
-\Gamma^{\omega}_{\mu\nu}\Gamma^{\sigma}_{\omega\rho}
\end{eqnarray}
The above action (\ref{nl1}) may be recast to the following  scalar-tensor form
using a pair of auxiliary (may be unphysical) scalar fields
 $\{\psi,\xi\}$:
\begin{eqnarray}
\label{nl2}
&&S=\int d^4 x \sqrt{-g}\left[
\frac{1}{2\kappa^2}\left\{R\left(1 + f(\psi)\right)
+ \xi\left(\Box\psi - R\right) \right\}
+ {\cal L}_{\rm matter}
\right] \\&& \nonumber
=\int d^4 x \sqrt{-g}\left[
\frac{1}{2\kappa^2}\left\{R\left(1 + f(\psi)\right)
 - \partial_\mu \xi \partial^\mu \psi - \xi R \right\}
+ {\cal L}_{\rm matter}
\right]
\ .
\end{eqnarray}
The equations of motion are supplemented by a set of Euler-Lagrange equations for $\xi$ in the following form:
\begin{eqnarray}
\frac{\delta S}{\delta \xi}=0,\ \ \Box\psi=R\label{eom1}
\end{eqnarray}
The above equation may be recast to the following form
$\psi=\Box^{-1}R$. If we  substitute this equation
 into (\ref{nl2}), we reobtain
(\ref{nl1}).
Equations of motion for metric tensor $g_{\mu\nu}$ is obtained from $\frac{\delta S}{\delta g_{\mu\nu}}=0$:
\begin{eqnarray}
\label{eom2}
0 &=& \frac{1}{2}g_{\mu\nu} \left\{R\left(1 + f(\psi) - \xi\right)
 - \partial_\rho \xi \partial^\rho \psi \right\}
 - R_{\mu\nu}\left(1 + f(\psi) - \xi\right) \\&& \nonumber
 + \frac{1}{2}\left(\partial_\mu \xi \partial_\nu \psi
 + \partial_\mu \psi \partial_\nu \xi \right)
 -\left(g_{\mu\nu}\Box - \nabla_\mu \nabla_\nu\right)\left( f(\psi) - \xi\right)
+ \kappa^2T_{\mu\nu}\ .
\end{eqnarray}
If we write the equation of motion for  $\psi$ we get the following equation of motion:
\begin{eqnarray}
\label{eom3}
0=\Box\xi+ f'(\psi) R\ .
\end{eqnarray}

Our aim here is to derive explicit forms of
(\ref{eom1},\ref{eom2},\ref{eom3}) for spherically symmetric static
configuration of compact, neutron, quark  \footnote{Quark (q) is a
fundamental fermion that has strong interactions}stars.

\section{Spherically symmetric model of compact stars}\label{Spherically symmetric model of compact stars}
We suppose there must be a broadly believable compact star in
static-spherically symmetric  coordinates given by system
$x^{\mu}=(ct,r,\theta,\varphi)$ in the following form:
\begin{eqnarray}
ds^2=c^2 e^{2\phi}dt^2-e^{2\lambda}dr^2-r^2(d\theta^2+\sin^2\theta d\varphi^2)\label{g}.
\end{eqnarray}
We also assume that the matter fields are same which have been used
in the interior parts and are in the comoving motion. The
appropriate energy-momentum tensor is expressed as
 $T_{\mu}^{\nu}=diag(\rho c^2,-p,-p,-p)$.  To keep the homogeneity, we assume that $\xi\equiv \xi(r),\ \ \psi\equiv\psi(r)$.
We insert (\ref{g}) in (\ref{eom2}) and  using the formula given in
Sec. (\ref{APPENDIX}), we obtain the diagonal components of
(\ref{eom2}) for  $(\mu,\nu)=(ct,ct)$, and $(\mu,\nu)=(r,r) $ are
given by the following differential equations:
\begin{eqnarray}
\texttt{tt}:\ \ \label{tt} && -\frac{1}{2}\xi'\phi'+(1+f(\phi)-\xi)(-\frac{2\lambda'}{r}+\frac{1-e^{2\lambda}}{r^2})-\Big(\psi''f_{\psi}
+\psi'^2f_{\psi\psi}-\xi''\Big)\\&&\nonumber-\Big(\frac{2}{r}-\lambda'\Big)(\psi'f_{\psi}-\xi')=\kappa^2\rho c^2 e^{2\lambda}.
\end{eqnarray}

\begin{eqnarray}
\texttt{rr}:\ \ (1+f(\phi)-\xi)(\frac{2\phi'}{r}+\frac{1-e^{2\lambda}}{r^2})+\frac{\xi'\phi'}{2}-(\frac{2}{r}+\phi')(\psi'f_{\psi}-\xi')=-\kappa^2e^{2\lambda}p
 \label{rr}.
\end{eqnarray}

\par
For scalar field $\psi$ we rewrite  (\ref{eom1}):
\begin{eqnarray}
&&\texttt{psi} :\ \ \psi''+\Big(\frac{2}{r}+\phi'-\lambda'\Big)\psi'-\phi''-\phi'^2+\phi'\lambda'-\frac{2}{r}(\phi'-\lambda')-\frac{1-e^{2\lambda}}{r^2}=0
 \label{psi}.
\end{eqnarray}
and similarly for (\ref{eom3}) we obtain:
\begin{eqnarray}
&&\texttt{xi}:\ \ \xi''+\Big(\frac{2}{r}+\phi'-\lambda'\Big)\xi'+2f_{\psi}\Big(\phi''+\phi'^2-\phi'\lambda'+2\frac{\phi'-\lambda'}{r}+\frac{1-e^{2\lambda}}{r^2}\Big)=0
\label{xi}.
\end{eqnarray}

The trace of the equation of motion (\ref{eom2}) provides another useful equation:
\begin{eqnarray}
&&R\left(1 + f(\psi) - \xi\right)-\partial_{\mu}\xi\partial^{\mu}\psi-3\Box (f(\psi)-\xi)=-\kappa^2(\rho c^2-p).
\end{eqnarray}
We are licensed to write the equation on to metric (\ref{g})
\begin{eqnarray}
&&\texttt{Trace}:\ \  \nonumber2\Big(\phi''+\phi'^2-\phi'\lambda'+\frac{2}{r}(\phi'-\lambda')+\frac{1-e^{2\lambda}}{r^2}\Big)(1+f(\psi)-\xi)-\psi'\xi'\\&&-3\Big[(\psi''f_{\psi}
+\psi'^2f_{\psi\psi}-\xi'')+(\frac{2}{r}+\phi'-\lambda')(\psi'f_{\psi}-\xi')\Big]=\kappa^2 e^{2\lambda}(\rho c^2-p)\label{trace}.
\end{eqnarray}

\section{Tolman-Oppenheimer-Volkoff Equations}\label{Tolman-Oppenheimer-Volkoff Equations}
The gravitational equations of motion must be supported by an
appropriate hydrostatic equation for the matter fields inside the
star. This equation is nothing just the familiar continuity equation
for energy-momentum tensor : $$\nabla_{\mu}T_{\nu}^{\mu}=0$$ If we
put $\nu=r$ in this equation with the metric (\ref{g}), we get:
\begin{eqnarray}
\frac{dp}{dr}=-(p+\rho c^2)\phi'\label{p}.
\end{eqnarray}
Now, keeping $\nu=t$ in mind, a question may arise that this
completely vanishes the possibility of constructing  new
hydrodynamic equation. We will note that equations
(\ref{tt},\ref{rr},\ref{psi},\ref{xi},\ref{trace}) are reduced to GR
by any $f(R)=R$ model in the action.
\par
We need to find the forms of the TOV equations. It is safe to
replace the metric part  $\lambda$ of the equations with a mass
function $M=M(r)$:
\begin{eqnarray}
e^{-2\lambda}=1-\frac{2GM}{c^2 r}\Longrightarrow \frac{G dM}{c^2 dr}=\frac{1}{2}\Big[1-e^{-2\lambda}(1-2r\lambda')\Big]\label{dM}.
\end{eqnarray}
Now, we must rewrite (\ref{tt},\ref{rr},\ref{psi},\ref{xi},\ref{trace}) in terms of $\{\frac{dp}{dr},\frac{dM}{dr},\rho\}$. Also, it is adequate to write equations in the dimensionless forms. It is desired to introduce the next set of the dimensionless parameters for field equations,
\begin{eqnarray}
M\to m M_{\odot},\ \ r\to r_{g} r,\ \ \rho\to \frac{\rho M_{\odot}}{r_{g}^3},\ \ p\to \frac{ p M_{\odot} c^2}{r_{g}^3}.
\end{eqnarray}
where $r_{g}=\frac{GM_{\odot}}{c^2}=1.47473 KM$ and $M_{\odot}$
Stands out for its  mass of the central Sun. Using these
dimensionless parameters we rewrite the continuity equation as the
following:
\begin{eqnarray}
&&\phi'=-\frac{p'}{(p+\rho )}\label{p2}.
\end{eqnarray}
  But (\ref{dM}) converts to the following:
\begin{eqnarray}
&&\lambda'=\frac{m}{r^2}\frac{1-\frac{r}{m}\frac{dm}{dr}}{\frac{2m}{r}-1}\label{dm}.
\end{eqnarray}
Furthermore, we have metric function as an equation in terms of the mass:
\begin{eqnarray}
&&e^{2\lambda}=\Big(1-\frac{2m}{r}\Big)^{-1},\ \ 1-e^{2\lambda}=\Big(1-\frac{r}{2m}\Big)^{-1}.
\end{eqnarray}
A  very widely expressed system, TOV equations, connecting metric to
matter is also found in the following forms:
\begin{eqnarray}
\texttt{tt}:\ \ &&\nonumber  \frac{p'\xi'}{2(p+\rho )}+(1+f(\phi)-\xi)(-\frac{2m}{r^3}\frac{1-\frac{r}{m}\frac{dm}{dr}}{\frac{2m}{r}-1}+\frac{1}{r^2\Big(1-\frac{r}{2m}\Big)})\\&&-\Big(\psi''f_{\psi}
+\psi'^2f_{\psi\psi}-\xi''\Big)-\Big(\frac{2}{r}-\frac{m}{r^2}\frac{1-\frac{r}{m}\frac{dm}{dr}}{\frac{2m}{r}-1}\Big)\Big(\psi'f_{\psi}-\xi'\Big)=\frac{8\pi\rho}{ 1-\frac{2m}{r}}
\label{tt2}.
\end{eqnarray}

\begin{eqnarray}
\texttt{rr}:\ \ \label{rr2} &&(1+f(\phi)-\xi)\Big(\frac{2p'}{r(p+\rho)}-\frac{1}{r^2\Big(1-\frac{r}{2m}\Big)}\Big)+\frac{\xi' p'}{2(p+\rho)}\\&&\nonumber+(\frac{2}{r}-\frac{p'}{p+\rho})(\psi'f_{\psi}-\xi')=\frac{8\pi p}{1-\frac{2m}{r}}.
\end{eqnarray}

\begin{eqnarray}
&&\texttt{psi} :\ \ \nonumber \psi''+\Big(\frac{2}{r}-\frac{p'}{p+\rho}-\frac{m}{r^2}\frac{1-\frac{r}{m}\frac{dm}{dr}}{\frac{2m}{r}-1}\Big)\psi'+\Big(\frac{p'}{p+\rho}\Big)'-\Big(\frac{p'}{p+\rho}\Big)^2\\&&-\frac{m}{r^2}\frac{p'}{p+\rho}\frac{1-\frac{r}{m}\frac{dm}{dr}}{\frac{2m}{r}-1}+\frac{2}{r}(\frac{p'}{p+\rho}+\frac{m}{r^2}\frac{1-\frac{r}{m}\frac{dm}{dr}}{\frac{2m}{r}-1})-\frac{1}{r^2\Big(1-\frac{r}{2m}\Big)}=0
\label{psi2}.
\end{eqnarray}
Similarly, an equation could find in(\ref{eom3}) :
\begin{eqnarray}\label{xi2}
&&\texttt{xi}:\ \  \xi''+\Big(\frac{2}{r}-\frac{p'}{p+\rho}-\frac{m}{r^2}\frac{1-\frac{r}{m}\frac{dm}{dr}}{\frac{2m}{r}-1}\Big)\xi'+2f_{\psi}\Big(-\Big(\frac{p'}{p+\rho}\Big)'\\&&\nonumber+\Big(\frac{p'}{p+\rho}\Big)^2+\frac{m}{r^2}\frac{p'}{p+\rho}\frac{1-\frac{r}{m}\frac{dm}{dr}}{\frac{2m}{r}-1}-\frac{2}{r}\Big(\frac{p'}{p+\rho}+\frac{m}{r^2}\frac{1-\frac{r}{m}\frac{dm}{dr}}{\frac{2m}{r}-1}\Big)+\frac{1}{r^2\Big(1-\frac{r}{2m}\Big)}\Big)=0.
\end{eqnarray}

\begin{eqnarray}\label{trace2}
&&\texttt{Trace}:\ \    2\Big(-\Big(\frac{p'}{p+\rho}\Big)'+\Big(\frac{p'}{p+\rho}\Big)^2+\frac{p'}{p+\rho}\frac{m}{r^2}\frac{1-\frac{r}{m}\frac{dm}{dr}}{\frac{2m}{r}-1}\\&&\nonumber-\frac{2}{r}(\frac{p'}{p+\rho}+\frac{m}{r^2}\frac{1-\frac{r}{m}\frac{dm}{dr}}{\frac{2m}{r}-1})\nonumber+\frac{1}{r^2\Big(1-\frac{r}{2m}\Big)}\Big)(1+f(\psi)-\xi)-\psi'\xi'\\&&-3\Big[(\psi''f_{\psi}
+\psi'^2f_{\psi\psi}-\xi'')\nonumber+(\frac{2}{r}-\frac{p'}{p+\rho}-\frac{m}{r^2}\frac{1-\frac{r}{m}\frac{dm}{dr}}{\frac{2m}{r}-1})(\psi'f_{\psi}-\xi')\Big]=\frac{8\pi(\rho -p)} {\Big(1-\frac{2m}{r}\Big)}.
\end{eqnarray}
Try adding up our model $f(R)$, we solve
(\ref{tt2},\ref{rr2},\ref{psi2},\ref{xi2},\ref{trace2}) numerically.
\section{An empirical model for the astrophysical objects}\label{Isotropic compact star}
In previous section we derived the full set of TOV equations for a
generic model of non-local $f(R)$ gravity. Our aim in this section
is to investigate a simple model of compact stars. By compact star,
we mean some relativistic massive objects  with tiny size and high
density\cite{book}. These astronomical objects have the mass of
order $M\sim M_{\odot}$ and the radius $R\sim 10 KM$. For three
types of the astronomical candidates  the metric functions
$\{\lambda,\phi\}$ were obtained  as simple quadratic functions of
radial coordinate $r$ \cite{KB}:
\begin{eqnarray}
&&2\lambda=Ar^2,\ \ 2\phi=Br^2+C,\label{ABC}
\end{eqnarray}
where $A$, $B$ and $C$ are physical constants to be evaluated  using
astronomical data given in the table I. In GR, compact stars with
these metric functions were studied. In modified gravity from
 $f(R)$ to the GB models \cite{Astashenok:2014gda}-\cite{Astashenok:2014nua}, there are some interesting features. In reference \cite{Astashenok:2014nua}, a model for neutron star constructed numerically for a class of viable models of $f(G)$ gravity. Our aim in this letter is to investigate physical properties of compact stars in this non local f(R) gravity (see \cite{Momeni:2014bua}
,\cite{Astashenok:2014nua} for some recently reported works). We
will study the dynamical stability, energy conditions and red shift
properties of an isotropic model of compact star. The model which we
address here is the non-local form of the one proposed by Starobinsky for inflation
\cite{Starobinsky}:
\begin{eqnarray}
f(\Box^{-1}R)=\Box^{-1}R+\alpha \Big(\Box^{-1}R\Big)^2.
\end{eqnarray}

Even the best known neutron stars often have great uncertainties in
their masses and radii. In fact no neutron star has a really precise
radius measurement, and it is rare to have even rough measurements
of both radius and mass for one star. The mass of strange star,
describes a fit with two minima of error, one is $0.8 M_{\odot}$,
the other being $1.8 M_{\odot}$ which is far more believable as it
is consistent with good measurements of other systems
\cite{arXiv:1201.5519}, and both mass values are highly dependent on
the details of this very complicated model, in contrast with the
simpler methods that produce good mass measurements in other
systems. This binary is complicated by x-ray heating of the primary,
among other things. Since even the work \cite{arXiv:1201.5519} say
that the $0.8 M_{\odot}$ measurement cannot be favored by their
analysis, it is unreasonable to take it as definitive. The $R\sim7
Km$ radius for SAX J1808 may extract from \cite{hep-ph/9905356}
which it seems to be still popular in the particle literature
although the astronomy literature shows it to be wrong, as in
\cite{astro-ph/0703287} and others. We still cannot figure out how
can we obtain the mass and radii for the star 4U 1820-30  (for a
recent work see for example \cite{arXiv:1305.3242} for a detailed
 and careful discussion, recent analysis including statistical and
(numerous) systematic sources of error).

\begin{table}[ht]
\caption{The values of parameters $\{A,B\}$. You may also use the
interpolated  data in
\cite{arXiv:1305.3242,hep-ph/9905356,astro-ph/0703287} which can be
accepted.}
\begin{center}
\begin{tabular}{|c|c|c|}
\hline { Strange star candidate }& \textbf{ $A(km ^{-2})$}& \textbf{$B(km
^{-2})$}
\\\hline  Her X-1& 0.006906276428 &
$0.004267364618$
\\\hline SAX J 1808.4-3658&  0.01823156974 &
$0.01488011569$
\\\hline 4U 1820-30&0.01090644119 &
$0.009880952381$
\\\hline
\end{tabular}
\end{center}
\end{table}


\begin{itemize}

\item {\bf  Exact solutions for auxiliary fields} $\Big[\psi,\xi\Big]$: the equations of motion (\ref{eom1},\ref{eom3}) are completely integrable :
\begin{eqnarray}
\psi(r)=C_1\int{\frac{g_{rr}dr}{\sqrt{-g}}}+\int_{0}^{r}{\sqrt{-g(y)}R(y)(r-y)dy}+C_2,\label{sol-psi}
\end{eqnarray}
here $R(y)$ is (\ref{R}). For $\xi$ using (\ref{eom3}) we obtain:
\begin{eqnarray}
\xi(r)=C_3\int{\frac{g_{rr}dr}{\sqrt{-g}}}-\int_{0}^{r}{\sqrt{-g(y)}(r-y)f_{\psi}(y)\Box\psi(y)dy}+C_4,\label{sol-xi}
\end{eqnarray}
Where $\Box\psi(y)$ was obtained from (\ref{Box}). However, it is
their hard work, integrating and  simplifying to achieve their
fields solutions (\ref{sol-psi}, \ref{sol-xi}). Fortunately, exact
solutions  (\ref{sol-psi}, \ref{sol-xi}) have given the graphically
monotonic forms. In Fig. (\ref{fig6}), we observe that $\psi(r)$
never falls into a linear plot, which remains the polynomial. So
auxiliary field $\psi(r)$ is to be the perfect polynomial plot
against the radius. An example Her X-1  plot of field $\chi(r)$ in
an interior area is shown in Fig. (\ref{fig7}). We conclude that
$\xi(r)$ is always decreasing  or remaining constant and never
increasing, so both of $\psi(r), \xi(r)$ are monotonic-increasing
(decreasing) functions. For Starobinsky model, this behavior is a
functional dependency of auxiliary fields on scalar curvature $R$ of
metric (\ref{ABC}). You may increase or decrease the Ricci scalar
$R$ (\ref{R}) simply by extending or decreasing the $\{A,B,C\}$, or
by increasing or decreasing the amount of fields $\psi,\xi$.
However, the increasing value of $R$ points to the growing
appearance of second derivative of the  scalar fields
$\{\Box\xi,\Box\psi\}$. At the same time both field(s) and growing
Ricci scalar are increasing (decreasing) behavior.


\begin{figure}
\begin{center}
\epsfxsize=10cm \epsfbox{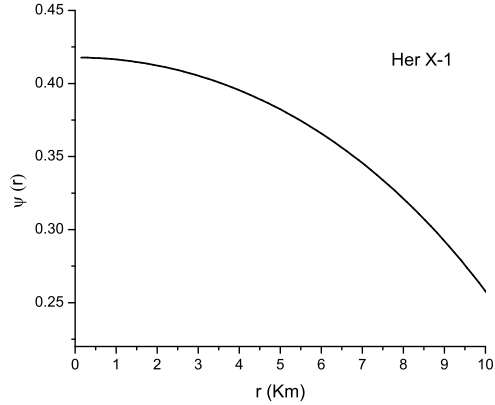} \caption{Numerical plot of the  $\psi(r)$ (\ref{sol-psi})}\label{fig6}
\end{center}
\end{figure}

\begin{figure}
\begin{center}
\epsfxsize=10cm \epsfbox{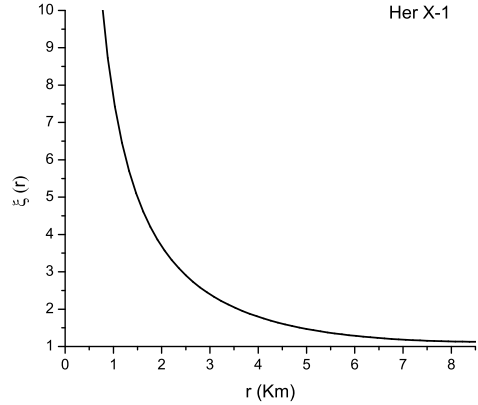} \caption{Numerical reconstructed  plot of $\xi(r)$ using (\ref{sol-xi})}\label{fig7}
\end{center}
\end{figure}

\item {\bf Stability-conditions}: with applying external radial perturbations at the inflow boundary, large radial structures develop naturally in the flow field due to sound effects. The velocity of sound drift, produced by spatially uniform perturbations is obtained by:
\begin{eqnarray}\nonumber
&&V_{rv}^{2}\equiv\frac{dp}{d\rho }\label{dp.drho}.
\end{eqnarray}
We plot (\ref{dp.drho}) in Fig. (\ref{fig2}). The probability
(velocity) for the radial perturbations will peak at order ten to
the one. Generally, however, there are instabilities with either
pressure  radial perturbations or with density to astrophysical
groups SAX J 1808.4-3658,Her X-1 and 4U 1820-30. One may increase or
decrease the $V_{rv}^{2}$ simply by extending or decreasing the
pressure, or by increasing or decreasing the amount of  density. For
SAX J 1808.4-3658, numerical analysis showed a significant similar
relationship between increasing proportions of sample SAX J
1808.4-3658, Her X-1 and 4U 1820-30. Initially, increasing the
pressure perturbation to the density perturbation increases the
velocity $V_{rv}^{2}$ of sample. The plot may have a singularity, or
the perturbation scheme may be inappropriate. But an instability
ought to come after the singular point. The earlier mentioned sample
SAX J 1808.4-3658 of astrophysical object is a such indicative case.
However, looking back we should have instabilities in  SAX J
1808.4-3658 a lot earlier than Her X-1 and 4U 1820-30. Thermal
instability of the stars with large perturbations is a function of
the empirical parameters $\{A,B,C\}$ of the sample.

\begin{figure}
\begin{center}
\epsfxsize=10cm \epsfbox{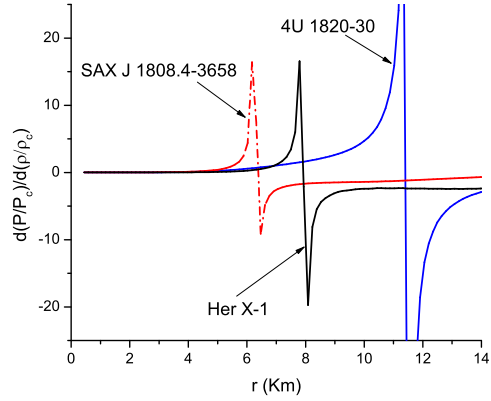} \caption{The probability (velocity) for the radial perturbations $\frac{d P/P_c}{d \rho/\rho_c}$}\label{fig2}
\end{center}
\end{figure}

\item {\bf Surface Redshift}: we've an increase in the wavelength of radiation emitted by a celestial body as a consequence of the gravitational field.
The gravitational redshift $z$ of thermal spectrum detected at infinity can be
computed as
\begin{eqnarray}
z=e^{-\phi}-1\label{z}.
\end{eqnarray}
We plot (\ref{z}) in Fig. (\ref{fig5}). It is observed that the largest redshift occurs for an emitter at the center of the star.
\begin{figure}
\begin{center}
\epsfxsize=10cm \epsfbox{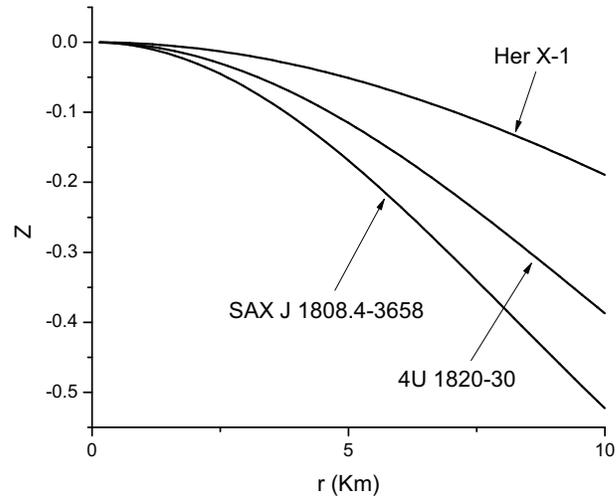} \caption{Redshift of photons (\ref{z}) emitted from the center of the star as a function of $r$ for three different candidates SAX J 1808.4-3658,Her X-1 and 4U 1820-30.}\label{fig5}
\end{center}
\end{figure}

\item  {\bf Energy conditions}: by looking at the field equations in modified gravities we are often able to  arrange a mutually acceptable energy density $\rho_{\text{eff}}$ and pressure $p_{\text{eff}}$. These energy conditions have altered our understanding of the range of conditions under which energy transfering onto a region occurs
  \cite{lobo,anzhong}:
\begin{eqnarray}
\text{NEC}&\Longleftrightarrow&\rho_{\text{eff}}+p_{\text{eff}}\geq0.\label{n1}\\
\text{WEC}&\Longleftrightarrow& \rho_{\text{eff}}\geq0\ \text{and}\ \rho_{\text{eff}}+p_{\text{eff}}\geq0.\label{n2}\\
\text{SEC}&\Longleftrightarrow& \rho_{\text{eff}}+3p_{\text{eff}}\geq0\ \text{and}\ \rho_{\text{eff}}+p_{\text{eff}}\geq0.\label{n3}\\
\text{DEC}&\Longleftrightarrow& \rho_{\text{eff}}\geq0\ \text{and}\ \rho_{\text{eff}}\pm p_{\text{eff}}\geq0.\label{n4}
\end{eqnarray}
 The essence of energy conditions is purely geometrical simple and, as Hawking said, self-evident
\cite{hawking}. However, we also believe that energy conditions in
their pure essence are effective. Fig. (\ref{fig1}) showing the
$\frac{\rho}{\rho_c}$ changed in order of  decrease of pressure
$\frac{p}{p_c}$ \footnote{In all of these graphs  ${\rho}_c= \frac{
M_{\odot}}{r_{g}^3}, p_c=\frac{M_{\odot} c^2}{r_{g}^3}$.}. The
resulting graph (\ref{fig1}) shows the rapid decrease in
$\frac{\rho}{{\rho}_c}$ in the three different candidates SAX J
1808.4-3658, Her X-1 and 4U 1820-30 and the more gradual decrease
through the 4U 1820-30 and SAX J 1808.4-3658. The resulting graph
(\ref{fig3}) shows the rapid decrease in scaled pressure
$\frac{p}{p_c}$ in three different candidates and the more gradual
decrease through the 4U 1820-30 and SAX J 1808.4-3658s. All the
pressures get vanished on the radius of star and none of the
pressures is remained at all. For Her X-1, the pressure is vanished
near the $r\sim 8.25$ and for SAX J 1808.4-3658, at $r\sim 6.5$ and
4U 1820-30 at $r\sim 11$. The precise distances for the vanished
pressures can be seen in the
\cite{arXiv:1305.3242,hep-ph/9905356,astro-ph/0703287}. The levels
of radius found numerically are comparable to those found by
astrophysical data
\cite{arXiv:1305.3242,hep-ph/9905356,astro-ph/0703287}. These two
methods clearly  produce comparable data. Information verified and
indexed by data from the (\ref{figp+r}, \ref{fig3p+r}, \ref{fig10})
would be easily cross-referenced in WEC and SEC. The satisfactions
of the all energy conditions will be independently verified
numerically. Then again we empirically verified that our non local
model for compact star was correct. The role of non locality in
compact stars was therefore verified in  saturation of the all
energy conditions. However, despite strong interest in their usage,
a lack of fundamental test data and verified structural non local
theory guidance is inhibiting uptake.

\begin{figure}
\begin{center}
\epsfxsize=10cm \epsfbox{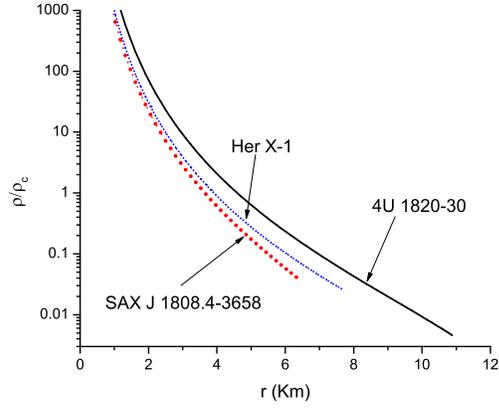} \caption{Scaled energy density  $\frac{\rho}{\rho_c}$ of the star as a function of $r$ for three different candidates SAX J 1808.4-3658,Her X-1 and 4U 1820-30. }\label{fig1}
\end{center}
\end{figure}

\begin{figure}
\begin{center}
\epsfxsize=10cm \epsfbox{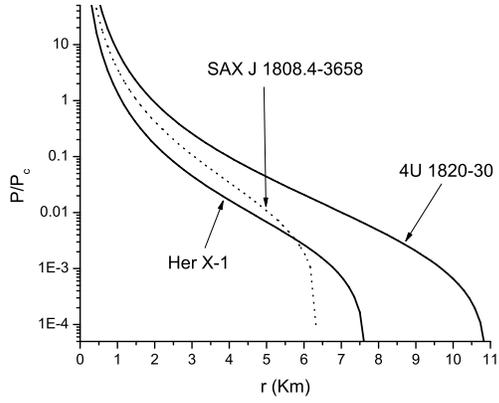} \caption{Scaled pressure  $\frac{p}{p_c}$ of the star as a function of $r$ for three different candidates SAX J 1808.4-3658,Her X-1 and 4U 1820-30. }\label{fig3}
\end{center}
\end{figure}

\begin{figure}
\begin{center}
\epsfxsize=10cm \epsfbox{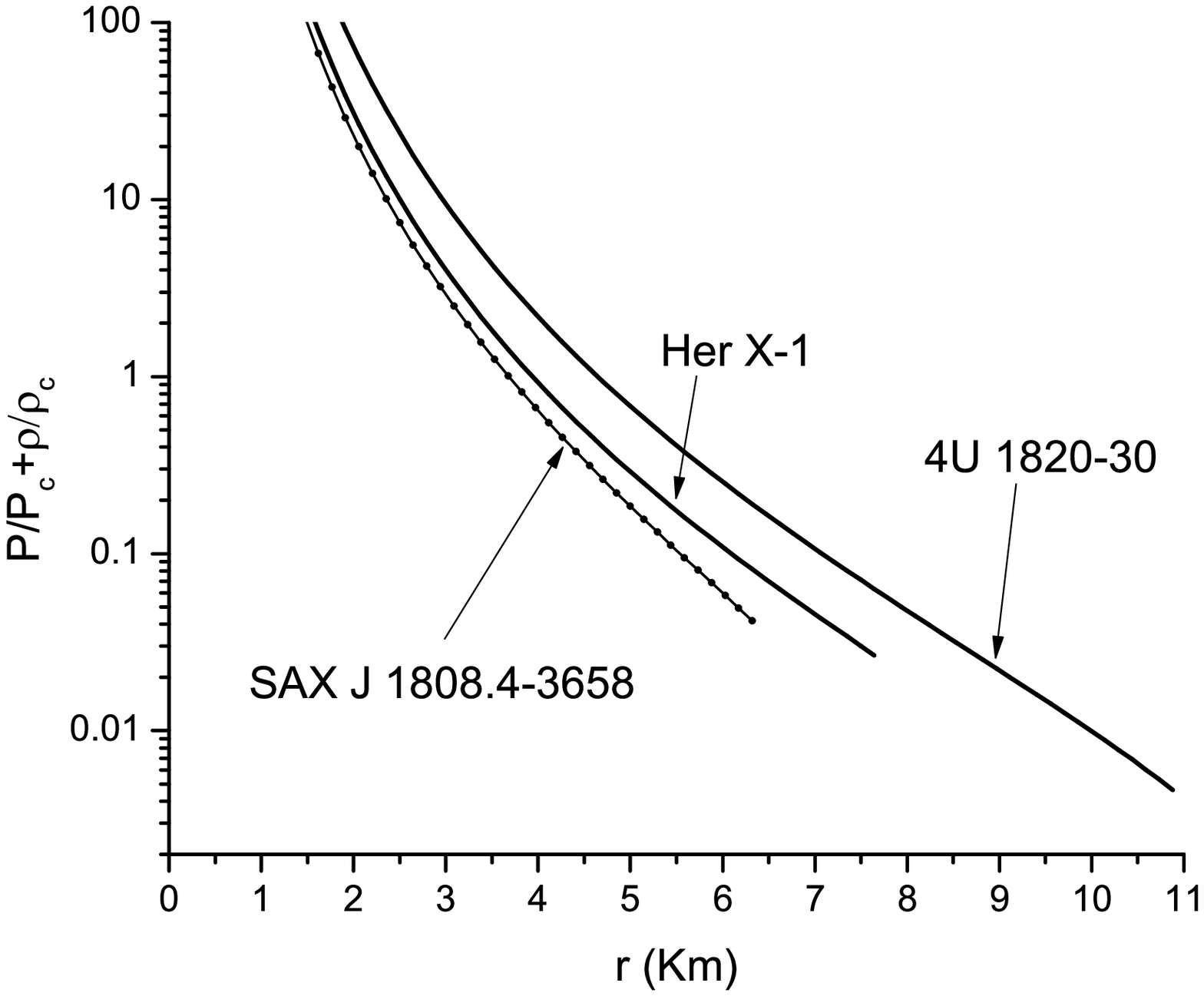} \caption{ Scaled $\frac{P}{P_c}+\frac{\rho}{\rho_c}$.}\label{figp+r}
\end{center}
\end{figure}

\begin{figure}
\begin{center}
\epsfxsize=10cm \epsfbox{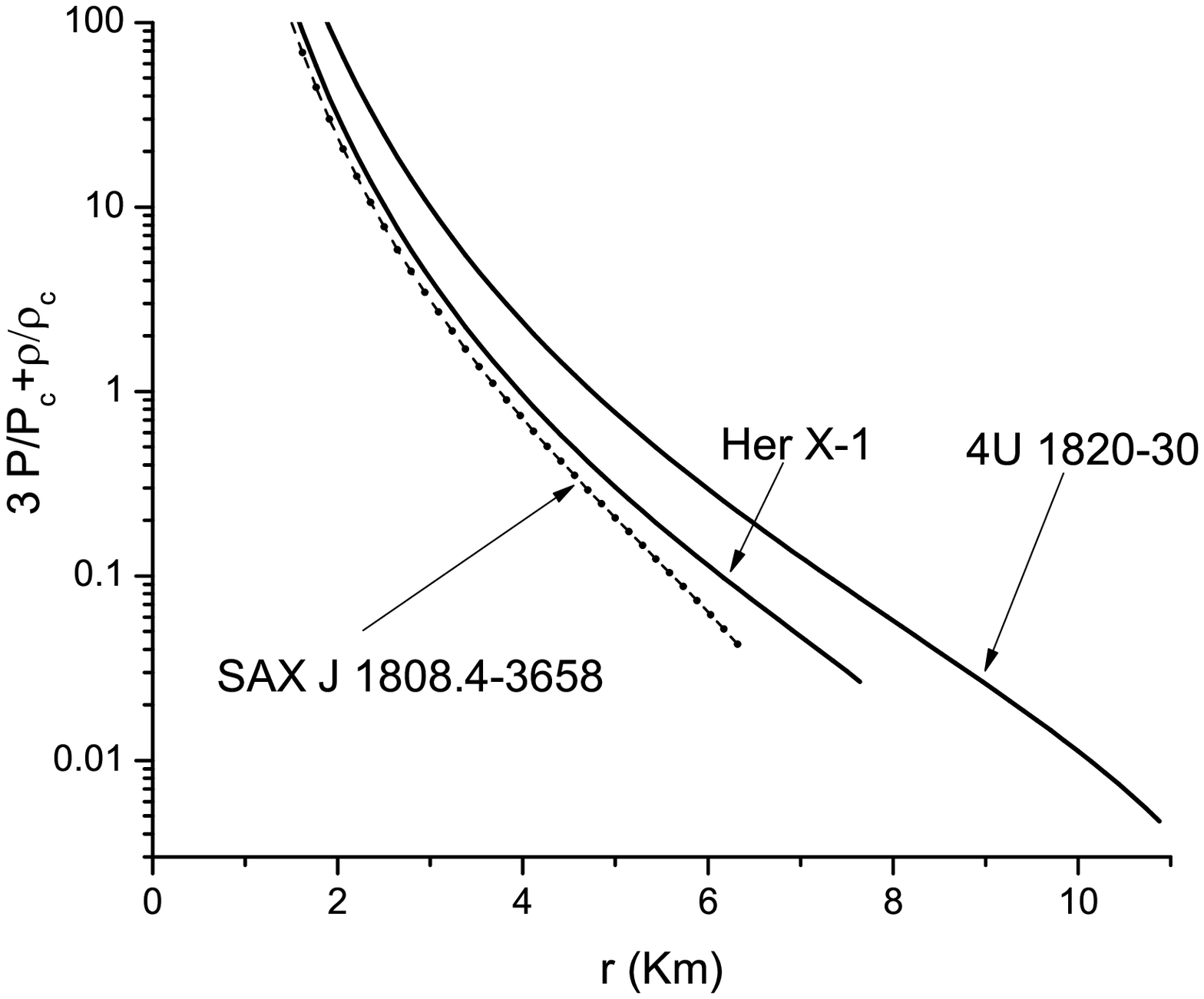} \caption{Scaled  $\frac{3P}{P_c}+\frac{\rho}{\rho_c}$}\label{fig3p+r}
\end{center}
\end{figure}

\begin{figure}
\begin{center}
\epsfxsize=10cm \epsfbox{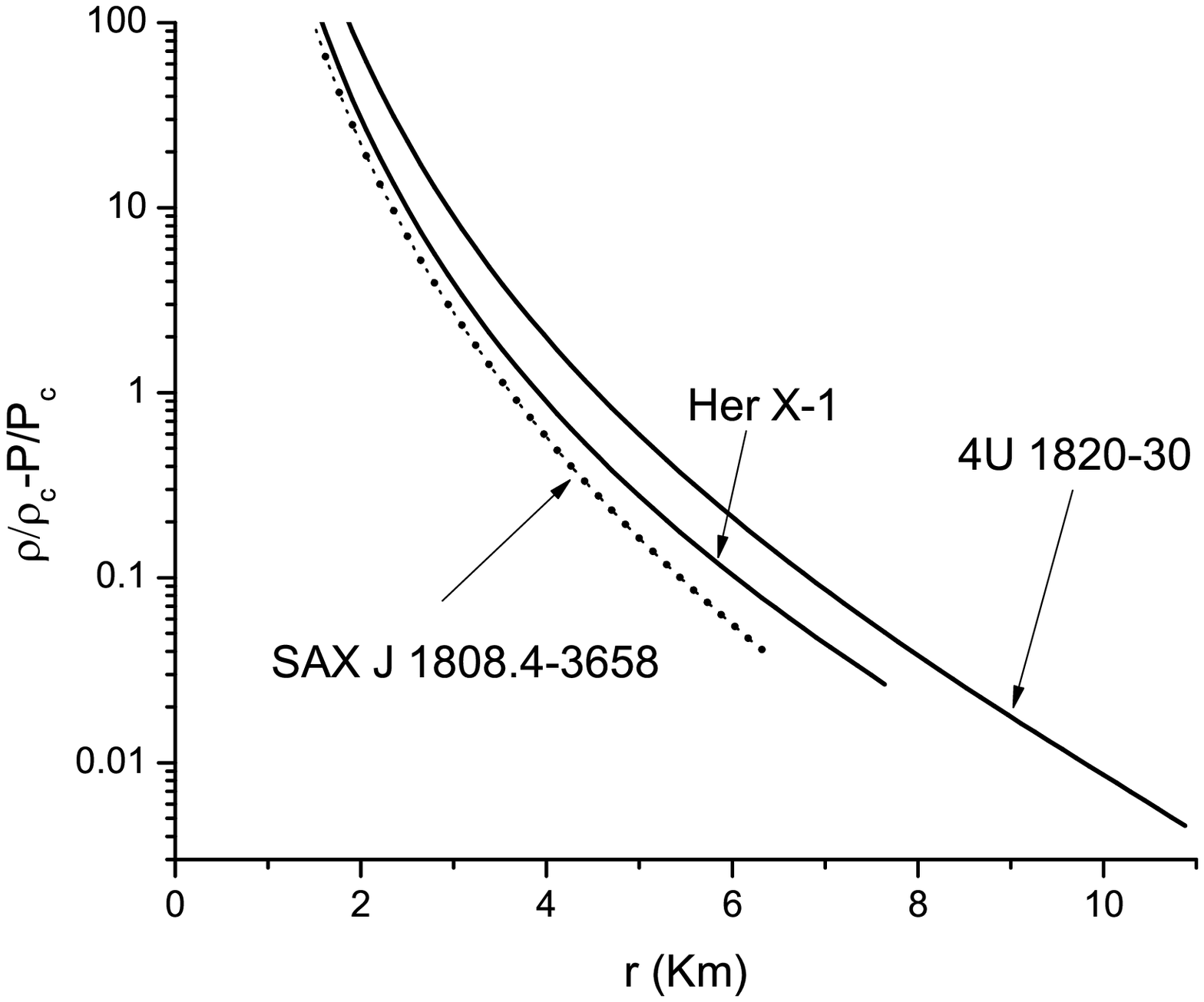} \caption{Numerical plot of the  $\frac{P}{P_c}-\frac{\rho}{\rho_c}$ }\label{fig10}
\end{center}
\end{figure}

\item{\bf  An empirical equation of state}:

Fig. (\ref{fig4}) shows the EoS   of a compact star calculated numerically. The results suggest that the appropriate form of EoS  is given by the following:
\begin{eqnarray}
\frac{P}{P_c}= a (1- e^{-b\frac{\rho}{\rho_c}}).\label{eos}
\end{eqnarray}
Tiny density increases the pressure size and gave value to the
hardenability of star. We claimed that increasing density $\rho$ "
will thwart the aspirations of future neutron star ". Quark stars
are also required to attend a generic EoS which emphasizes linear
relation $p=A\rho+B$. However, if the density is decreased, you
would be required to attend the quark star EoS at the non local
$f(R)$ model. Normally, someone contemplating these low densities
$\rho\ll \rho_c$ would attend quark star EoS. All parameters are
shown in Table 2. Some parameters offer a selection from a sample to
which you give an option. For some objects parameters, for example
4U 1820-30, high pressure effects are stronger than pressure effects
in other samples. Finally, the EoS parameters must also be
consistent with a linearity in the quark EoS which leads to the
observed samples. These parameters are tested in both density and
pressure initial conditions. We mention here that the EoS
(\ref{eos}) can be addressed as a generalized exponential Virial EoS
\cite{Kenneth}:
\begin{eqnarray}
\label{Kenneth}&&\frac{p}{\rho k_B T}=\exp\Big(\Sigma_{m=2}^{\infty}K_m \rho^{m-1}\Big).
\end{eqnarray}
where the coefficients $K_m$ and the virial coefficients $B_n$ are
related, $k_B$ is the Boltzmann constant and $T$ temperature. There
are significant differences in the EoS of (\ref{eos}) versus
exponential Virial EoS (\ref{Kenneth}). These differences have been
categorized in terms of pressure of the background $p_b\sim p_c$.
For instance, for anomalous differences all the normal gases with
exponential Virial EoS  have an $p_b$ of zero. These differences
arise because one or more compact objects in our case has a $p_b$
value or indeed should not be omit. However, it is debatable whether
these structural differences make this $EoS$ any easier for the
astrophysical purposes to investigate the thermodynamic. The
differences between (\ref{eos}) and exponential Virial EoS
(\ref{Kenneth}) are:  (\ref{eos}) is made from the gravitational
field, whereas  exponential Virial EoS (\ref{Kenneth}) is made from
the  high-density fluid. In conclusion, it is of utmost importance
to reiterate the differences between the  (\ref{eos}) EoS and the
exponential Virial EoS (\ref{Kenneth}) .

\begin{figure}
\begin{center}
\epsfxsize=10cm \epsfbox{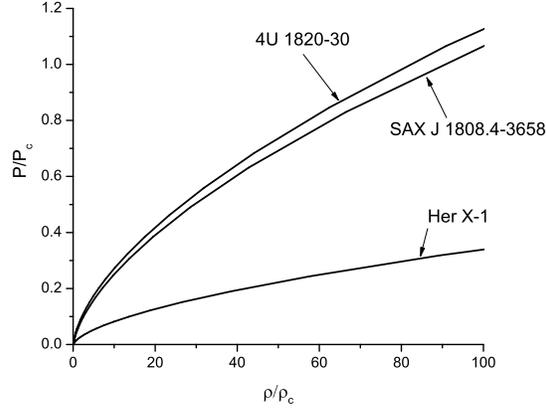} \caption{Empirically computed quation of state $\frac{P}{P_c}= a (1- e^{-b\frac{\rho}{\rho_c}})$. Here $\sigma_a,\sigma_b$ are standard errors for data fitting.  }\label{fig4}
\end{center}
\end{figure}

\begin{table}[h]
\caption{The parameters of EOS for three different stars : $\frac{P}{P_c}= a (1- e^{-b\frac{\rho}{\rho_c}})$. }
\begin{center}
\begin{tabular}{|c|c|c|c|c|}
\hline
        Astrophysical  strange star candidate          & a       & $\sigma_a$ & b        & $\sigma_b$ \\ \hline
Her X-1           & 0.45526 & 0.0198     & 0.0152   & 0.00124    \\ \hline
SAX J 1808.4-3658 & 1.2242  & 0.0563     & 0.0191   & 0.00162    \\ \hline
4U 1820-30        & 1.1255  & 0.0408     & 0.0243   & 0.00163    \\ \hline
\end{tabular}
\end{center}
\end{table}

\end{itemize}

\section{Summary and conclusion}\label{Summary and conclusion}
Einstein gravity is a gauge theory of gravity. It should be modified
to have more effective predictions and implications for recently
observational data. One of the most accepted modifications of
Einstein gravity is $f(R)$ gravity and its extensions. It is assumed
that we gain more information about gravity if we replace $R$ by an
arbitrary function $f(R)$. Several cosmological aspects of this type
of modified gravity have been investigated in literature. Specially
the late and early time evolution. Non-local terms, induced by
quantum effects can be considered as non-local higher order
corrections to Einstein gravity. It is reasonable to consider both
scenarios in a same context, as non-local $f(R)$ gravity, a scenario
which we studied in this letter. We derived the equations of motion
for this non-local theory using a pair of auxiliary scalar fields.
As a motivated idea, we studied stellar structure using the modified
forms of TOV equations. We obtained the set of equations of motion
for a star in non-local form of $f(R)$ gravity. At the same time the
TOV equations were recast and some reconstruction took place in the
system, when a non-local correction was inserted. It is asserted
here that the dynamic can adequately describe, explain or understand
such a naive relationship from the perspective of  compact stars.
The full conditional solutions are evaluated and inverted
numerically to obtain exact forms of the compact stars Her X-1,SAX J
1808.4-3658 and 4U 1820-30 for model of $f(\Box^{-1}R)=\Box^{-1}R+\alpha \Big(\Box^{-1}R\Big)^2$.  The
program solves the differential equations  numerically using
adaptive Gaussian quadrature. An ascription of correctness is
supposed to be an empirical equations of state  $\frac{P}{P_c}= a
(1- e^{-b\frac{\rho}{\rho_c}})$ for star which is informative in so
far as it excludes an alternative non local approach to compact star
formation. The differences between (\ref{eos}) and exponential
Virial EoS (\ref{Kenneth}) are:  (\ref{eos}) is made from the
gravitational field, whereas  exponential Virial EoS (\ref{Kenneth})
is made from the  high-density fluid. This model is most suited for
astrophysical observation. A theoretical perspective proposed by us,
TOV equations for non-local $f(R)$ theory, is helpful for
understanding these non-local effects.

\section*{Acknowledgments}

This work has been supported financially under project   "\emph{evolution of black holes and wormholes in a modified theory of gravity}" in program  named
 "\emph{some problems of the nonlinear theory of gravitational and strong interactions and their cosmological applications}". The author gratefully acknowledges valuable suggestions of the reviewer to improve our work.


\section{Appendices}

In this appendix we present different geometrical quantities which
have been used in this letter. For metric (\ref{g}) the following
nonzero components of the symmetric connection are obtained:
\begin{eqnarray}
\Gamma_{12}^1=\phi',\ \ \Gamma_{11}^2=\phi' e^{2\phi-2\lambda},\ \ \Gamma_{22}^{2}=\lambda',\ \ \Gamma_{23}^2=-re^{-2\lambda},\\
\Gamma_{44}^2=-r\sin^2\theta e^{-2\lambda},\ \ \Gamma_{23}^3=\frac{1}{r},\ \ \Gamma_{44}^3=-\sin\theta\cos\theta, \ \ \Gamma_{24}^4=\frac{1}{r},\ \ \Gamma_{34}^4=\cot\theta.
\end{eqnarray}
So, the nonzero $(tt),(rr)$ components of the Ricci  tensor read:
\begin{eqnarray}
R_{tt}&=&e^{2\phi-2\lambda}(-\phi''-\phi'^2+\phi'\lambda'-\frac{2\phi'}{r}),\\
R_{rr}&=&\phi''+\phi'^2-\phi'\lambda'-\frac{2\lambda'}{r}.
\end{eqnarray}
The Ricci scalar is as the following:
\begin{eqnarray}
R=-2e^{-2\lambda}\Big(\phi''+\phi'^2-\phi'\lambda'
+\frac{2(\phi'-\lambda')}{r}+\frac{1-e^{2\lambda}}{r^2}\Big)\label{R}.
\end{eqnarray}
The operator $\Box$  is given by:
\begin{eqnarray}
\Box A(r)=-e^{-2\lambda}\Big(A''+\Big(\frac{2}{r}+\phi'-\lambda'\Big)A'\Big)\label{Box}.
\end{eqnarray}


\end{document}